\documentclass[twocolumn,amsmath,amssymb,aps]{revtex4-1}

\usepackage{graphicx}
\usepackage{dcolumn}
\usepackage{bm}
\usepackage{color}
\usepackage{amsmath}

\newcommand{\be}{\begin{equation}}
\newcommand{\ee}{\end{equation}}

\newcommand{\bea}{\begin{eqnarray}}
\newcommand{\eea}{\end{eqnarray}}

\def\eq#1{Eq.~(\ref{#1})}
\def\eqs#1#2{Eqs.~(\ref{#1},\ref{#2})}

\newcommand\pO{\phi_0}
\newcommand\dvO{\delta v_0}
\newcommand\dvOp{\delta v_0'}
\newcommand\dd{\mathrm{d}}
\newcommand\DD{\mathcal{D}}
\newcommand\s{\sigma_s}

\newcommand\ew{\epsilon_w}
\newcommand\eo{\epsilon_0}

\newcommand\kv{\kappa_v}

\newcommand\qc{\Lambda_c}

\newcommand\kb{\kappa_b}
\newcommand\kc{\kappa_c}

\newcommand\lb{\ell_{\rm B}}
\newcommand\lc{\ell_c}

\renewcommand\ln{\mathrm{ln}}
\newcommand\bl{\bar{\lambda}}
\newcommand{\nb}{\eta_b}
\newcommand{\pmf}{\Phi}

\newcommand{\at}[1]{\arctan\left( #1 \right)}

\renewcommand{\vec}[1]{\mathbf{#1}}
\renewcommand{\vr}{\vec{r}}

\renewcommand{\DH}{Debye-H\"uckel }

\renewcommand{\v}{v}

\renewcommand{\AA}{\text{\normalfont\r{A}}}

\begin{document}

\title{A variational approach to the liquid-vapor phase transition for hardcore ions in the bulk and in nanopores}

\author{Bastien Loubet}
\affiliation{Laboratoire de Physique Th\'eorique, IRSAMC, Universit\'e de Toulouse, CNRS, UPS, France}
\author{Manoel Manghi}
\affiliation{Laboratoire de Physique Th\'eorique, IRSAMC, Universit\'e de Toulouse, CNRS, UPS, France}
\author{John Palmeri}
\affiliation{Laboratoire Charles Coulomb (L2C), UMR 5221 CNRS-Universit\'e de Montpellier,
F-34095 Montpellier, France}

\date{\today}
\begin{abstract}

We employ a field-theoretical variational approach to study the behavior of ionic solutions in the grand canonical ensemble. To describe properly the hardcore interactions between ions, we use a cutoff in Fourier space for the electrostatic contribution of the grand potential and the Carnahan-Starling equation of state with a modified chemical potential for the pressure one. We first calibrate our method by comparing its predictions at room temperature with Monte Carlo results for excess chemical potential and energy. We then validate our approach in the bulk phase by describing the classical  ``ionic liquid-vapor'' phase transition induced by ionic correlations at low temperature, before applying it to electrolytes at room temperature confined to nanopores embedded in a low dielectric medium and coupled to an external reservoir of ions. The ionic concentration in the nanopore is then correctly described from very low bulk concentrations, where dielectric exclusion shifts the  transition up to room temperature for sufficiently tight nanopores, to high concentrations where hardcore interactions dominate which, as expected, modify only slightly this ionic ``capillary evaporation''.
\end{abstract}

\pacs{Valid PACS appear here}

\maketitle

\section{Introduction}

Charged hard spheres have been shown to exhibit a bulk phase transition between an ionic ``liquid'' state and an ionic ``vapor'' state for low enough density and temperature~\cite{Stell1976,Fisher1993,Ding1996,Diehl1997}. This transition is entirely due to positional electrostatic correlations between oppositely charged ions and cannot therefore be described by mean-field theory, whose contribution to the free energy vanishes by charge electroneutrality (see the review~\cite{Levin2002}).
In the ionic vapor phase oppositely charged ions tend to form neutral but polar groups which tend to interact loosely with each others. In the ionic liquid phase, the ions do not form such pairs and are directly screened by the others. Although this bulk phase transition for common mineral salts is predicted to occur in aqueous solutions at unphysical low temperatures ($T<100\ \mathrm{K}$) and is therefore not observable in experiments with classical electrolytes, some exotic ones do show this phase transition around ambient temperature~\cite{Weingartner1992}, especially in low dielectric constant solvents. By evoking the law of corresponding states the same model of charged hard spheres in a dielectric continuum can be used to model electrolytes, salts in low dielectric constant solvents,  molten salts (of great current interest for their potential technological applications), and perhaps also metallic fluids.

We have recently proposed a field theoretic variational approach that allows us, by going beyond mean-field theory, to study the effect of confinement and dielectric discontinuities on the ionic liquid-vapor phase transition~\cite{Buyukdagli2010a,Buyukdagli2010,Buyukdagli2011}. The method used previously did not take into account ionic size and therefore led to anomalies when extended to high electrolytes concentrations. It is, however, important to be able to go reliably to high electrolyte concentrations, because this same theoretical framework can be used to calculate the effect of confinement and dielectric exclusion on the ionic transport coefficients currently being measured for well characterized single nanopores~\cite{Siria2014,Balme2015}. At very high electrolyte concentration one expects the transport coefficients to tend toward their bulk values and the measured deviations as the concentration is lowered could provide valuable insight into the transport mechanisms.

Previous theoretical works included hardcore interactions coupled with Coulomb interactions. Modified Poisson-Boltzmann approaches, for example, have been developed by introducing an explicit expression for the free energy contribution of the hardcore interactions, which is based either on lattice gas calculations~\cite{Borukhov1997}, or on the Carnahan-Starling pressure~\cite{Lue1999,Maggs2016}. Netz and Orland~\cite{Netz1999}, working in the canonical ensemble, introduced an ultra-violet cut-off, $\Lambda_c$, in the Debye electrostatic free-energy calculation in order to take into account the excluded volume effect, following Brilliantov who did it for the one-component plasma~\cite{Brilliantov1998}. They obtained corrections to the well known Debye-H\"uckel volumetric free-energy density valid in the limit $\Lambda_c\to \infty$~\cite{Debye1923}
\begin{equation}
f_{\rm DH}= -k_{\rm B}T\frac{\kappa_b^3}{12\pi}
\label{fDH}
\end{equation}
where $k_{\rm B}T$ is the thermal energy and $\kappa_b$ the Debye screening parameter. Although these corrections take into account in an approximate way the effect of finite ion-size on the electrostatic interactions, they do not account for direct hardcore interactions. Using a field-theoretical model which includes them, Moreira and Netz~\cite{Moreira2002} derived the first coefficients of the virial expansion of a non-symmetric electrolyte, valid for low densities. In Ref.~\cite{Coalson1995} a Yukawa potential was introduced to model short range repulsive interactions and a variational approach to a similar model was studied in~\cite{Buyukdagli2011b}. One of the most successful approaches in describing the Monte Carlo (MC) results~\cite{Valleau1980,Valleau1991,Abbas2009} for the bulk phase transition is the physically motivated but \textit{ad hoc} model developed by Fisher and Levin~\cite{Fisher1996}, where a free energy with an explicit ion pairing term, or Bjerrum association, was constructed (see also~\cite{Yeh1996}). Recently 
Giera \textit{et al.}~\cite{Giera2015} performed molecular dynamics simulations of electric double layers and compared successfully the measured capacitance to the Carnahan-Starling mean-field calculation.\\

In this paper we extend our previously developed variational approach~\cite{Buyukdagli2011} by including the Carnahan-Starling pressure contribution in the variational grand potential. We  show that it is necessary to include both a hardcore regularization (\textit{via} a Fourier space wave vector cut-off) in the electrostatic part of the grand potential and the direct hardcore interactions in order to recover the correct behavior for the chemical potential and internal energy, computed using Monte Carlo simulations~\cite{Valleau1980}.
We subsequently apply our model to a fluid confined in a nanopore connected to reservoirs of ions and explore the partition coefficient of the ions in the pore and the modification of the phase transition (induced by the dielectric exclusion) due to the hardcore interactions.
In Section~\ref{sec:var} the general field theoretic approach, including the electrostatic and hardcore interactions, as well as the variational scheme used in this paper (and developed in Appendix~\ref{appA}) are exposed.
In Section~\ref{sec:bulk} we compute the bulk grand potential by introducing a wave vector cutoff in its electrostatic contribution and using the Carnahan-Starling pressure to express the hardcore one.
We then investigate the low temperature phase transition in the bulk in order to validate our model. We apply it in Section~\ref{sec:nano} to the case of an electrolyte confined in a neutral cylindrical nanopore inside a low dielectric medium, where we obtain a corrected behavior for the partition coefficient due to hardcore interactions. We also investigate the phase transition for the case of a weakly charged nanopore.
Finally our conclusions and perspectives are presented in Section~\ref{sec:concl}.

\section{General field theoretic variational model}
\label{sec:var}

We consider an electrolyte made of $N_\nu$ ions of type $\nu=1,\ldots,p$ in solution in water. We work in the grand canonical ensemble where the temperature $T$, the volume $V$ are fixed together with the ionic chemical potentials $\mu_\nu$ or equivalently their fugacity
\begin{equation}
\lambda_\nu=\frac{\exp(\mu_\nu)}{V_\nu}
\end{equation}
where {$V_\nu$ is a reference volume that does not enter into the final results for physical quantities}, and we express all the energies in units of $\beta^{-1}=k_\mathrm{B} T$.  Ions $\alpha$ and $\gamma$ interact through the electrostatic potential and a short range repulsive potential $U_{\alpha \gamma}$ and are submitted to an external potential, $u_\alpha (\vec{r})$, {acting} on particle $\alpha$.

We formulate the grand partition function in two steps. First we consider the hardcore potential alone by artificially setting the ion charges to zero.
The hardcore grand partition function is therefore
\begin{multline}
\label{grandpotHC}
 \Xi_{\rm hc} =\sum_{N_1 = 0}^\infty \cdots \sum_{N_p = 0}^\infty \prod_{\nu=1}^{p} \frac{\lambda_\nu^{N_\nu}}{N_\nu !}
  \int \prod_{\nu=1}^{p} \prod_{j=1}^{N_\nu} \dd \vec{r}_{\nu j} g(\vec{r}_{\nu j})  \\
\times \exp\left[-\frac{1}{2} \sum_{\alpha,\gamma=1}^p \int_{\vec{r},\vec{r}'} \hat{c}_\alpha(\vec{r}) U_{\alpha\gamma}(\vec{r}-\vec{r}') \hat{c}_\gamma(\vec{r}') \right. \\
\left. + \frac{1}{2} \sum_{\alpha=1}^p N_\alpha U_{\alpha\alpha}(0){-} \sum_{\alpha=1}^p \int_{\vec{r} }\hat{c}_\alpha(\vec{r}) u_\alpha(\vec{r})\right]
\end{multline}
where $\vec{r}_{\nu j}$ is the position of the $j$th particle of type $\nu$, and $g(\vec{r})$ is a function that models the possible restriction of the volume accessible to the particles. The density operators, $\hat{c}_\alpha$, are:
\begin{equation}
 \hat{c}_\alpha(\vec{r}) = \sum_{j=1}^{N_\alpha} \delta ( \vec{r} - \vec{r}_{\alpha j})
\end{equation}
We obtain the corresponding field theoretic formulation by applying a Hubbard-Stratonovitch transformation~\cite{Negele1992}:
\begin{multline}
 \exp\left(-\frac{1}{2} \sum_{\alpha\gamma} \int_{\vec{r},\vec{r}'} \hat{c}_\alpha(\vec{r}) U_{\alpha\gamma} \hat{c}_\gamma(\vec{r}') \right) = \\
 \frac{1}{Z_U}\int \prod_\nu \DD \psi_\nu \, \exp \left( -H_{\rm hc}[\psi_\gamma]  + i \sum_\alpha \int_{\vec{r}} \psi_\alpha(\vec{r}) \hat{c}_\alpha(\vec{r}) \right)
\end{multline}
where $Z_U = -\frac12\,\mathrm{tr}\,\ln(U)$ and the hardcore Hamiltonian is
\be
H_{\rm hc}[\psi_\gamma] =\frac12 \sum_{\alpha\gamma} \int_{\vec{r},\vec{r}'} \psi_\alpha(\vec{r}) U^{-1}_{\alpha\gamma}(\vec{r}-\vec{r}') \psi_\gamma(\vec{r}')
\ee
After summation over the number of particles $N_\nu$, \eq{grandpotHC} becomes:
\begin{multline}
\label{grandpotHCfield}
 \Xi_{\rm hc}{\left[ \mu_\alpha - u_\alpha(\vec{r}) \right]} = \frac{1}{Z_U}\int \prod_\nu \DD \psi_\nu  \exp \left(-H_{\rm hc}[\psi_\gamma]\right) \\ \times \exp \left[\sum_\alpha \frac{e^{\frac{1}{2} U_{\alpha\alpha}(0)}}{V_\alpha} \int_{\vec{r}} g(\vec{r}) e^{i \psi_\alpha(\vec{r}) {+ \mu_\alpha - u_\alpha(\vec{r})}} \right]
\end{multline}

Next we introduce the electrostatic contribution to the grand partition function. Ions $\alpha$ and $\gamma$ interact with the electrostatic energy $q_\alpha q_\beta v_{\rm c}(\vec{r})$ where $q_\alpha$ is the ion valency and the Coulomb interaction, $v_{\rm c}$, is
\be
 v_{\rm c}(\vec{r}) = \frac{\lb}{\left|\vec{r}\right|}
 \label{Coulomb}
\ee
 with $\lb = \beta e^2 / (4\pi \eo \ew)$  the Bjerrum length and $\ew\approx 78$ is the permittivity of water ($e$ is the quantum of charge).
The corresponding electrostatic contribution to the Hamiltonian is therefore
\begin{equation}
 H_{\rm el} = \frac{1}{2} \int_{\vec{r},\vec{r}'} \hat{\rho}(\vec{r}) v_{\rm c}(\vec{r}-\vec{r}') \hat{\rho}(\vec{r}') - \frac{1}{2} \sum_{\alpha=1}^p N_\alpha q_\alpha^2 v_{\rm c}(0)\label{Hel}
\end{equation}
where $\hat{\rho}$ is the charge density operator
\begin{equation}
 \hat{\rho}(\vec{r}) = \sum_{\alpha=1}^p q_\alpha \hat{c}_\alpha (\vec{r})
\end{equation}
The {factor of 1/2 in first  term of the rhs. of \eq{Hel} avoids  double counting and the second term subtracts the self-interactions}. The charge density leads to the introduction of an additional field $\phi$ in the field-theoretic description, in a fashion similar to the introduction of the fields $\psi_\alpha$ above.

{As shown by Netz \textit{et al.}~\cite{Netz2000,Moreira2002} the grand partition function of the system including both the hardcore and the electrostatic interactions can be written as a double functional integral over the fluctuating fields,
$\psi_\gamma$ and $\phi$. Although the ensuing problem cannot be treated exactly, we show in the Appendix~\ref{appA} that a powerful approximate variational method can still be employed to handle the electrostatic part.
As in previous variational approaches without hardcore interactions~\cite{Netz2003,Curtis2005,Hatlo2008,Hatlo2008a}, we  treat the electrostatic part by choosing a variational Hamiltonian $H_0$ with a general Gaussian form:
\be
 H_0[\phi] =\frac12 \int_{\vec{r},\vec{r}'} \left[  \phi(\vec{r}) - i \phi_0(\vec{r}) \right] v_0^{-1}(\vec{r},\vec{r}') \left[  \phi(\vec{r}') - i \phi_0(\vec{r}') \right]
 \label{HOdef}
\ee
where the variational functions are the average field $\phi_0(\vec{r})$ and the Gaussian kernel $v_0^{-1}(\vec{r},\vec{r}')$.  The variational grand-potential accounting for both electrostatic and hardcore interactions is given by (see Appendix~\ref{appA})
 \bea
 \Omega_{v} &=& - \ln (\Xi_{\rm el,hc}^{\rm v}) = \Omega_0+
 \langle  H_{\rm c}[\phi(\vec{r})] - H_0[\phi(\vec{r})]  \rangle_0 \nonumber \\
 &+&
 \int_\vec{r}  \rho_{\rm e}(\vec{r})\phi_0(\vec{r})
 +\Omega_{\rm hc} [\mu_\gamma - u_\gamma^0(\vec{r})],
 \label{varpot}
 \eea
 where
 $\Omega_0=-\frac12\mathrm{tr}\, \ln(v_0/v_{\rm c})$, the expectation value is evaluated with the  variational Gaussian Hamiltonian $H_0[\phi]$,
 $\rho_{\rm e}(\vec{r})$ is the external fixed charge density (in units of $e$),
\be\label{Hc0}
H_{\rm c}[\phi]=\int \mathrm{d}\vec{r} \frac{\epsilon(\vec{r})}{2\beta e^2}  [\nabla\phi(\vr)]^2,
\ee
is the Coulomb Hamiltonian with a spatially dependent permittivity $\epsilon (\vr)$ and
\bea
\Omega_{\rm hc} [\mu_\gamma - u_\gamma^0(\vec{r})] &=& -\ln \left( \Xi_{\rm hc} [\mu_\gamma - u_\gamma^0(\vec{r})] \right) \nonumber\\
&=& -V P_{\rm hc} [\mu_\gamma - u_\gamma^0(\vec{r})],
\label{omegahc}
\eea
 is the exact hardcore grand-potential (or minus the normalized osmotic  pressure times volume) in an external field equal to
 \be
 u_\gamma^0(\vec{r}) = q_\gamma \phi_0(\vec{r}) +  \frac{1}{2} q^2_\gamma
 \left[ v_0(\vr, \vr) - v_{\rm c}(0)\right].
 \ee
We will see below that the last term in $ u_\gamma^0(\vec{r})$ is the excess chemical potential  and therefore $\mu_\gamma - u_\gamma^0(\vec{r})$ is simply equal to the ideal gas contribution, 
\be
\mu_\gamma^{\rm id} = \ln (V_\gamma c_\gamma), 
\ee
where $c_\gamma$ is the concentration of ion $\gamma$. The variational grand-potential, $\Omega_{\rm v} $,  is an upper bound to the exact grand-potential $\Omega_{\rm el,hc} = - \ln (\Xi_{\rm el,hc})$.}

\eq{varpot} is then a sum of an electrostatic contribution minus an osmotic pressure one, $V P_{\rm hc}$, created by a hardcore fluid with a modified fugacity that is expressed as a function of the variational fields $\phi_0(\vr)$ and $\v_0(\vr,\vr)$.

For simplicity we work in the following with a restricted variational method by choosing the inverse kernel $v_0$ to be the solution of the inhomogeneous variational \DH equation~\cite{Buyukdagli2011},
\begin{equation}
 [-\nabla( \epsilon(\vr) \nabla) + \epsilon(\vr) \kappa^2(\vr)]v_0(\vec{r},\vec{r}') = \beta e^2 \delta(\vec{r} - \vec{r}'),
\end{equation}
and the average potential and  variational inverse screening length to be constant:
\begin{equation}
 \phi(\vec{r}) = \phi_0 \quad \mathrm{and} \quad \kappa(\vec{r}) = \kv g(\vec{r})
\end{equation}
\eq{varpot} is then minimized with respect to the restricted variational parameters $\kv$ and $\phi_0$.

 \section{Bulk electrolyte with hardcore interactions}
 \label{sec:bulk}

\subsection{Excess chemical potential}

The variational grand potential \eq{varpot} without the hardcore contribution, i.e. without the last term, has been computed  in Ref.~\cite{Buyukdagli2011}. The bulk contribution, i.e. with $\rho_e=0$, reads per unit volume:
\begin{equation}
   \label{wbnohc}
  w_{v} \equiv \frac{\Omega_{v}}{V} =
  -\sum_i \lambda_i  e^{\frac{q_i^2}{2} \lb \kv}
  + \frac{\kv^3}{24 \pi}
\end{equation}
where $\kappa_v$ is the only variational parameter because the mean-field $\phi_0$ vanishes in the bulk due to overall
charge neutrality (the index $i$ denotes the type of ion). The first term is minus the osmotic pressure of an ideal solution of ions and the second one is the usual Debye-H\"uckel term, which in the canonical ensemble leads to \eq{fDH}. Note that, in \eq{wbnohc}, $w_v\to-\infty$  when $\kv\to\infty$. Physically this divergence means that the most stable state of the system without hardcore interaction is a state of infinite concentration of neutral ionic pairs on top of each other.
Although the variational theory without hardcore interactions is not rigorously well defined, for sufficiently low salt concentrations the (metastable) minimum of \eq{wbnohc} with respect to $\kv$ does yield the standard \DH inverse length $\kb$, where
 \begin{equation}
 \label{DHkappanohc}
  \kb^2 = 4\pi \lb \sum_i q_i^2 c_i ,
 \end{equation}
which is obtained from the grand potential per unit volume $w_v$:
 \begin{equation}
  c_i = - \lambda_i \frac{\partial w_v }{\partial \lambda_i}
  \label{defc}
 \end{equation}
The first term in \eq{wbnohc}, with $\kv$ replaced by $\kb$, is therefore minus the osmotic pressure of an ideal solution, $\sum_ic_i$. Indeed, within the variational approach the excess electrostatic chemical potential is given by
\bea
\mu_{{\rm el},i}^{\rm ex} &\equiv& \ln(\lambda_i/c_i) \nonumber \\
&=& \frac{1}{2} q^2_i \left[ v_0(\vr, \vr) - v_{\rm c}(0)\right] = -\frac{q_i^2}{2} \lb \kb, \label{muexnohc}
\eea
where in the bulk system the inverse kernel takes on the Debye-H\"uckel form:
\begin{equation}
 v_0(\vec{r},\vec{r}') =\lb \frac{e^{-\kb\left|\vec{r} - \vec{r}' \right|}}{\left|\vec{r} - \vec{r}' \right|}.
\end{equation}

Note that $\mu_{{\rm el},i}^{\rm ex}\to -\infty$ when $\kb \to -\infty$. It costs less and less energy to add an ion into the system as the concentration increases which can again be interpreted as the collapse of the ions on each other in the absence of hardcore repulsion.
The second term in \eq{wbnohc} is the \DH electrostatic correlation contribution to the grand potential. Although this form is correct for low concentrations, it is not reliable for high concentrations, where the hardcore interaction should dominate.

\begin{figure}[t]
\includegraphics[width=\columnwidth]{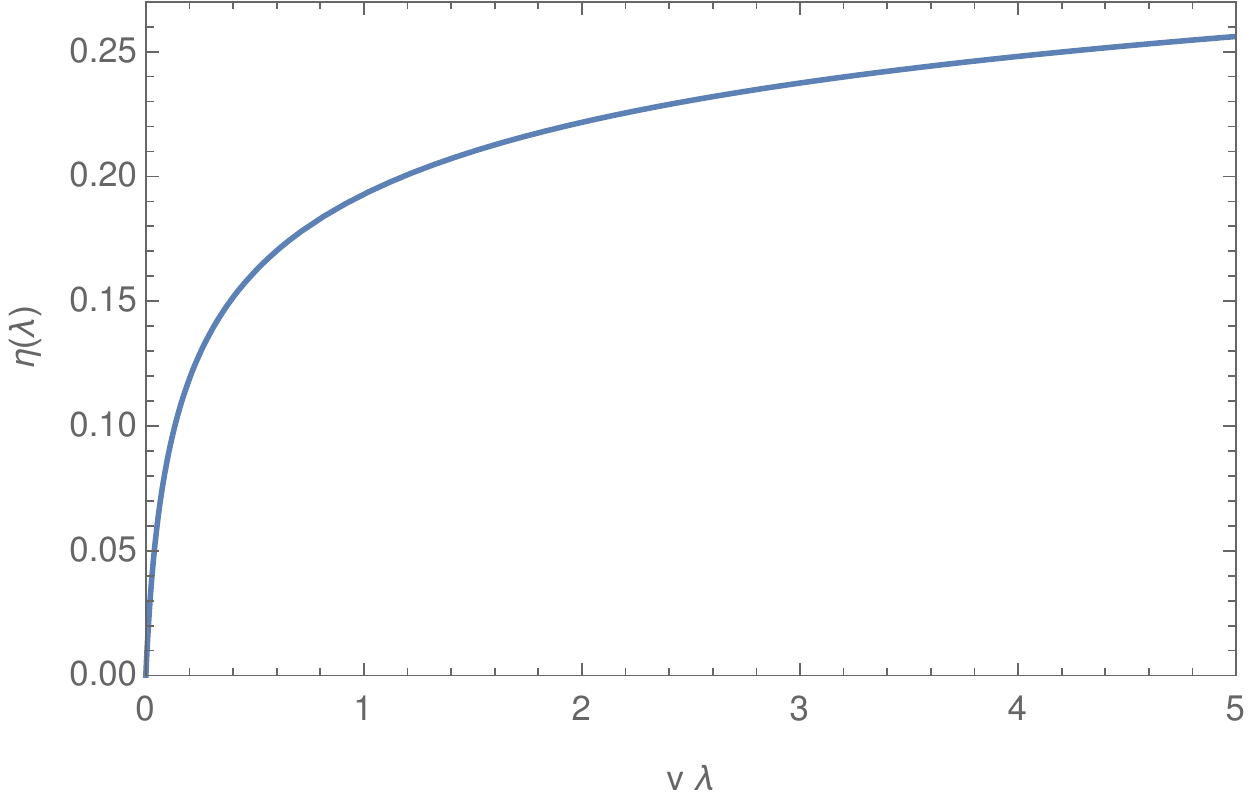}
\caption{The packing fraction $\eta(\lambda)$ as a function of $\v \lambda$. The function is zero at $\lambda = 0$ and slowly saturates to $1$ as $\lambda \rightarrow \infty$.}\label{fig:eta}
\end{figure}

To take into account the hardcore repulsion, we  use the variational prescription delineated above, and replace the first term of \eq{wbnohc} by minus the (grand canonical) pressure of a hardcore liquid, $-P_{\rm hc}$. To implement the variational method we  choose to approximate this hardcore pressure by the well known Carnahan-Starling form, which is almost quantitatively exact for neutral liquids up to freezing densities~\cite{Carnahan1969,Lue1999,Hansen2007}.
Explicitly, we use the following replacement:
\begin{equation}
 \sum_i \lambda_i  e^{ -\mu_{{\rm el},i}^{\rm ex} }  \rightarrow  P_{\rm hc}\left(\sum_i \lambda_i e^{ -\mu_{{\rm el},i}^{\rm ex} }\right)
\end{equation}
 where
 \begin{equation}
 \label{CSpress}
  P_{\rm hc}(\bl) = \frac{1}{\v} \eta(\bl) \frac{1 + \eta(\bl) + \eta(\bl)^2 - \eta(\bl)^3}{(1-\eta(\bl))^3}
 \end{equation}
  is the Carnahan-Starling result with  $\v = \pi d^3/6$  the excluded volume and $d$ the particle diameter.
  In the pure hardcore system there are no Coulombic interactions and the packing fraction is:
  \begin{equation}
  \eta = \v \sum_i c_i
  \label{packfrac}
  \end{equation}
with $0\leq \eta \leq 1$. For sake of clarity, we decide to consider here only the case where anions and cations have the same diameter (the so-called Restricted Primitive Model). The case of different diameters will be treated in a future work. In our grand canonical approach a relation has to be given between the fugacity and the packing fraction. The total concentration of hard sphere particles is $c = \sum_i c_i = \bl {\partial P_{\rm hc}(\bl )}/{\partial \bl}$ and therefore the packing fraction, \eq{packfrac} is related to the pressure given in \eq{CSpress} by:
\begin{equation}
  \label{eqdiffeta}
  \eta(\bl) =  \v \bl \frac{\partial P_{\rm hc}(\bl )}{\partial \bl},
  \end{equation}
which can be integrated and rewritten as a self-consistent equation for $\eta(\bl)$:
 \begin{equation}
  \label{relfugacity}
   \bl = \frac{1}{\v} \eta(\bl) e^{\mu_{\rm hc}^{\rm ex}(\eta(\bl))},
  \end{equation}
where the Carnahan-Starling excess chemical potential due to hardcore interactions,
  \begin{equation}
  \label{muexhc}
   \mu_{\rm hc}^{\rm ex}(\eta) = \frac{\eta-3}{(\eta-1)^3}-3,
  \end{equation}
has been deduced using the definition $\bl = c \ e^{\mu_{\rm hc}^{\rm ex}}$. Note that the passage to \eq{relfugacity} is equivalent to integrating the thermodynamic relation ${\partial P_{\rm hc}(c)}/{\partial c} =  c {\partial \mu_{\rm hc}(c)}/{\partial c}$ (at constant $T$ and $V$) and that given $\bl$, \eq{relfugacity} has to be solved numerically, although $\eta(\bl)$ as a function of $\v \bl$ can easily be plotted parametrically (see Figure \ref{fig:eta}).

To introduce in an approximate way the hardcore interaction directly into the electrostatic part (beyond the introduction of the hardcore pressure), we compute the \DH term in the grand potential by introducing a cut-off in Fourier space, $0\leq|{\bf q}|\leq \Lambda_c$.
This is similar to what has been done in Ref.~\cite{Netz1999}, except that we work in the grand canonical ensemble and the appropriate thermodynamic potential is the grand potential.
By integrating in Fourier space the first two terms of \eq{varpot}, we obtain
\begin{multline}
\delta w_{vb} = \left( \Omega_0+ \langle  H_{\rm c}[\phi(\vec{r})] - H_0[\phi(\vec{r})]  \rangle_0 \right)/V  = \\
 -\frac{\kv^2 \qc }{12 \pi^2} + \frac{\kv^3}{12 \pi^2} \at{\frac{\qc}{\kv}} +\frac{\qc^3}{12 \pi^2} \ln \left(1 + \frac{\kv^2}{\qc^2}\right)
\end{multline}
Although for small $\kv$ we recover the $\kv^3/24 \pi$ behavior, the role of the cut-off parameter $\qc$ is to avoid the unphysical divergence of the excess electrostatic chemical potential, \eq{muexnohc} for large $\kv$.

The resulting variational grand potential for the bulk phase is:
\be
\label{bulkgrdpot}
w_{vb}(\kv) = - P_{\rm hc} \left( \sum_i \lambda_i e^{\frac{q_i^2}{\pi} \kv \lb \at{\frac{\qc}{\kv} }} \right) + \delta w_{vb}
\ee
For large $\kv$ Eq.\eqref{bulkgrdpot} now implies that $w_{vb}(\kv) \sim \ln(\kv/\qc)$ remains finite and positive and the ions can no longer collapse on each other.

Minimizing \eq{bulkgrdpot} with respect to $\kv$ and using \eq{relfugacity} leads to the following variational {equation for the solution} $\kappa_b$:
\bea
\kappa_b^2 = &\frac{4\pi\lb}{v} \frac{\sum_i q_i^2 \lambda_i e^{\frac{q_i^2}{\pi} \kb \lb \arctan\left(\frac{\qc}{\kv} \right)}}{\sum_i \lambda_i e^{\frac{q_i^2}{\pi} \kb \lb \arctan\left(\frac{\qc}{\kb} \right)}}\nonumber\\
&\times \eta\left(\sum_i \lambda_i e^{\frac{q_i^2}{\pi} \kb \lb \arctan\left(\frac{\qc}{\kb} \right)}\right)
\label{vareq}
\eea
where the function $\eta(\bar \lambda)$ is defined in \eq{relfugacity}.

Computing the concentration  $c_i$ of each {ion using} \eq{defc} leads, after some {rearrangement}, to the same \DH result for the inverse screening length, \eq{DHkappanohc}.
Using \eq{vareq} leads again to \eq{packfrac}, which means that $\eta$ is still the system packing fraction (even in the presence of combined hardcore and electrostatic interactions). The variational result for the bulk phase grand potential is then $w_{b} = w_{vb}(\kb)$ and the total excess ionic chemical potential is the sum of a regularized electrostatic  and  hardcore contributions, \eq{muexhc},
\bea
\label{muex}
 \mu_i^{\rm ex} &=& \mu_{{\rm el},i}^{\rm ex} +  \mu_{{\rm hc}}^{\rm ex}
 \nonumber \\
 &=&  -\frac{q_i^2}{\pi} \kb \lb \at{\frac{\qc}{\kb}}+ \left[\frac{\eta-3}{(\eta-1)^3}-3 \right].
\eea

The approach adopted here leads to an excess chemical potential that is simply the direct sum of a regulated electrostatic part and a pure hardcore part. The electrostatic part $\mu_{{\rm el},i}^{\rm ex}$ saturates to a finite value for $\kb\to\infty$ and to \eq{muexnohc} for $\kb\to 0$.
Hence  for large $\kb$, $\mu_{\rm ex}$ is dominated by the hardcore contribution \eqref{muexhc}, which diverges (in the Carnahan-Starling approach) at $\eta = 1$. We emphasize that in order to get an appropriate physical result for electrolytes two ingredients are necessary: the electrostatic part has to be regularized  (e.g., by introducing a Fourier space cut-off into the inverse kernel) and the hardcore interactions have to be included explicitly. This is in contrast to a previous approaches adopted for one component plasmas~\cite{Netz1999}, where one or the other feature, but not both, was added.
\begin{figure}[t]
\includegraphics[width=\columnwidth]{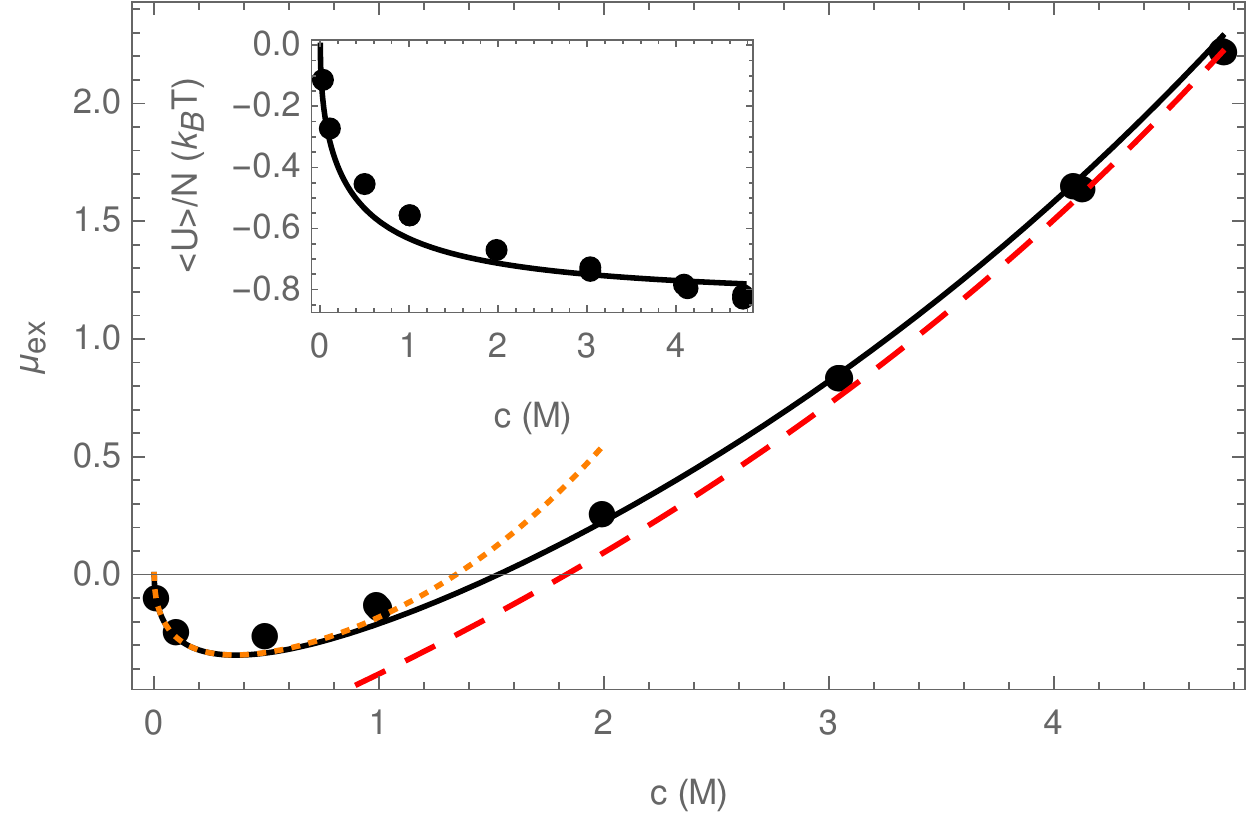}
\caption{The excess chemical potential in the bulk phase as a function of the ionic concentration. The solid line is the result of our model, \eq{loggamma}, fitted to the MC data by Valleau \textit{et al.}~\cite{Valleau1980} shown as circles. The orange dotted and red dashed lines are curves that capture the asymptotic behavior of the model for low and  large concentrations respectively. \textit{Inset}: Average excess internal energy per ion (circles: MC data of~\cite{Valleau1980}, line: our model without additional fitting parameter).}
\label{fig:gammabulk}
\end{figure}

The mean excess chemical potential for a simple salt, defined as
\be
\mu_\pm^{\rm ex} = \frac{\nu_{+} \mu_{+}^{\rm ex} + \nu_{-} \mu_{-}^{\rm ex}}{\nu_{+} +  \nu_{-}},
\ee
where $\nu_\pm$ are the stoichiometric coefficients (electroneutrality implies $\nu_{+} q_{+} = \nu_{-}|q_{-}|$), can be obtained from the  excess ionic chemical potentials, $\mu_i^{\rm ex}$.

The method proposed here also provides a variational foundation to canonical ensemble approaches to ion fluids (with and without hardcore interactions) formulated previously:  either by integrating \eq{muex} or performing a Legendre transformation on the grand potential leads to a bulk Helmholtz Free energy, $F_b$,
\be
f_{b} = \frac{F_{b}}{V} = w_{b} + \sum_i  c_i \mu_i = f_{b}^{\rm el} + f_{b}^{\rm hc},
\ee
that is the direct sum of a regularized electrostatic part and a hardcore contribution (for the Fourier space cut-off approach adopted here for the electrostatic contribution, $f_{b}^{\rm el}$ is identical to the result obtained in Ref.~\cite{Netz1999}).

We now investigate further the simplest case, that of a symmetrical $q-q$ electrolyte with $\lambda_+=\lambda_-=\lambda$ and $q_+=-q_-=q$. The excess chemical potential \eq{muex} simplifies to:
\bea
\mu_\pm^{\rm ex} (c_b) &=& -\frac{(2\lb q^2)^{3/2}}{\sqrt{\pi}} \sqrt{c_b} \at{\frac{\qc}{\sqrt{8\pi\lb q^2c_b}}} \nonumber \\
&+& \left[ \frac{2vc_b - 3}{(2vc_b-1)^3} - 3 \right] \label{loggamma}
\eea
where we have used $\nb = 2 \v c_b$ and $\kappa_b^2=8\pi\lb q^2c_b$ (we assume that $\nu_\pm =1$ and therefore $c_b$ is the salt concentration and the concentration of both anions and cations and $2 c_b$ is the total concentration of hardcore particles).
The first term on the right hand side of \eq{loggamma} is the contribution from the cut-off regularized electrostatic to the chemical potential.
This term always decreases when the concentration increases and tends to $- q^2 \lb \qc/\pi $ for large concentrations.
The second term, which is the hardcore contribution to the chemical potential, increases when the concentration increases.
The excess chemical potential, \eq{loggamma}, is plotted in Figure~\ref{fig:gammabulk} vs the bulk concentration together with the Monte Carlo (MC) simulation results obtained at room temperature by Valleau \textit{et al.}~\cite{Valleau1980}, shown as circles. 
We have fitted \eq{loggamma} to these data using $\qc$ as a parameter and $v = \pi d^3/6$ with $d = 4.25\ \AA$ the hardcore particle diameter used in the MC simulations~\cite{Valleau1980}. The fit is shown as a solid line in Figure \ref{fig:gammabulk} and yields $\qc = 3.73\ \mathrm{nm^{-1}}$. The agreement is very good. The average excess internal energy per ion is shown as circles in the inset and the curve correspond to our model without any additional parameter. The fitted value of the cut-off is $\sim 1/d$, an expected physically reasonable result (when compared with the usual DH approach~\cite{McQuarrie2000}) lends credence to our approach. 

For small $c_b$ the excess chemical potential reduces to
\begin{equation}
\mu_\pm^{\rm ex}  \approx  -\frac{(2\lb q^2)^{3/2}}{\sqrt{\pi}} \sqrt{c_b}  + 8\left(\frac{q^4\lb^2}{\qc} +v\right)c_b,  
\end{equation}
and this approximation is shown as a dotted line in Figure \ref{fig:gammabulk}.
It is clear that the large electrostatic term in the second virial contribution to the excess chemical potential obtained  perturbatively by Netz and Orland~\cite{Netz1999} for electrolytes at low concentrations has its origin in the regularization of the modified Coulomb interaction at short distances arising from hardcore interactions (an effect that for small enough ions typically dominates over the direct hardcore contribution). Note that expanding the first term on the right hand side of \eq{loggamma} for low $c_b$ leads to an alternating series and keeping the next term in $c_b^{3/2}$ without the hardcore term $v c_b$ would lead to a worse result.

For large concentrations, $c_b\lesssim c_{\max}=(2v)^{-1}$, the behavior is dominated by the hardcore contribution shifted by a constant:
\begin{multline}
 \mu_\pm^{\rm ex} \approx  \frac{2vc_b - 3}{(2vc_b-1)^3} - 3 \\ -\frac{2(\lb q^2)^{3/2}}{\sqrt{\pi v}} \at{\frac{\qc\sqrt{v}}{\sqrt{4\pi\lb q^2}}}
\end{multline}
This curve is shown in Figure~\ref{fig:gammabulk} (red dashed curve).
In conclusion, the relatively simple method presented here enables us to account for the thermodynamic properties of the Restrictive Primitive Model from low concentrations up to saturation, nearly rivaling the accuracy of much more sophisticated liquid theory methods (see. e.g., Ref.~\cite{McQuarrie2000}).

\subsection{Bulk ionic liquid-vapor phase transition}
\begin{figure}[t]
\includegraphics[width=\columnwidth]{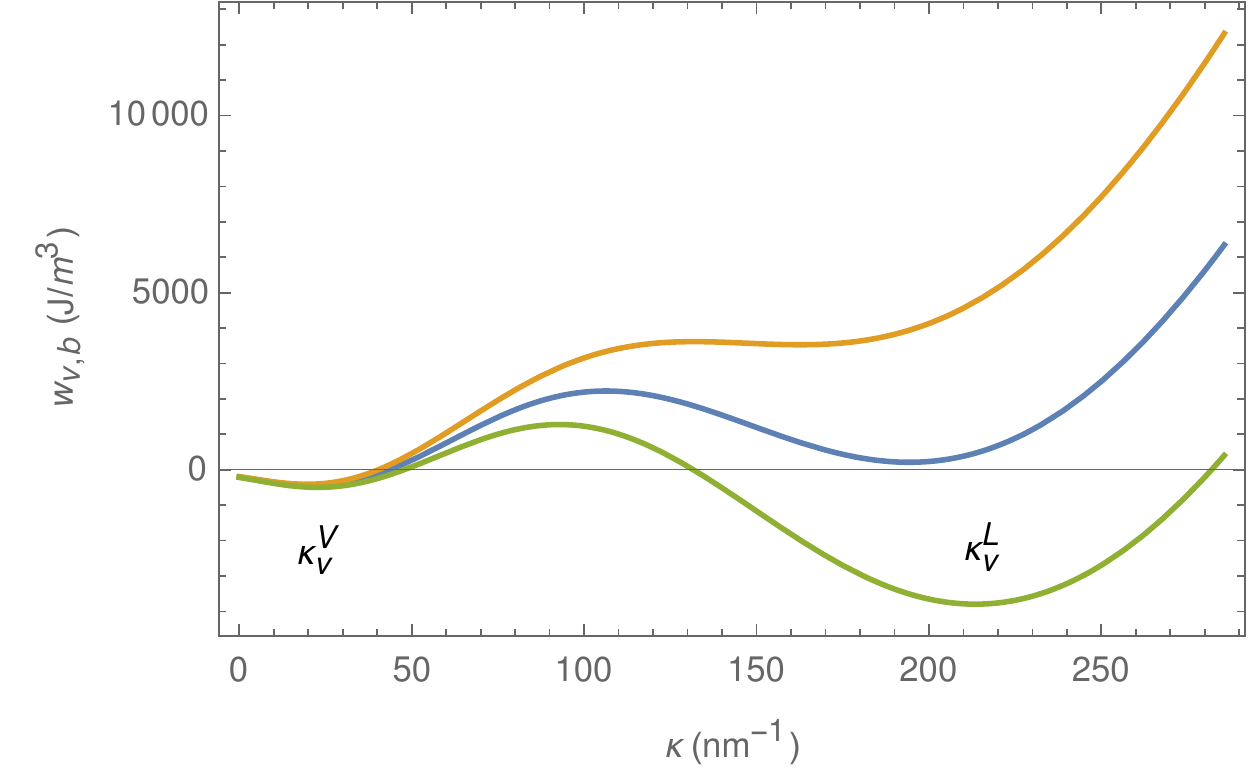}
\caption{The grand potential per unit volume $w_v$ as a function of the variational parameter $\kb$ for $\lb^3 \lambda=0.085$, 0.09 and 0.095 (from top to bottom). For each curve the temperature is kept fixed and the chemical potential increases from top to bottom. The low $\kv$ minima correspond to the vapor phase while the large $\kv$ ones correspond to the liquid phase.}
\label{fig:wvb}
\end{figure}

The low temperature bulk ionic phase transition governed by ion-ion correlations and discussed in the Introduction is a natural consequence of our model. In our variational scheme, this phase transition is characterized by the presence of two minima of $w_{v}(\kv)$ at fixed fugacity $\lambda$ and temperature $T$. Hence phase coexistence corresponds to multiple solutions to the variational equation, $\partial w_{vb}(\kv)/\partial\kv = 0$, for sufficiently low temperature. We illustrate this point in Figure~\ref{fig:wvb} where, by increasing $\lambda$ at low $T$ a second minimum of $w_{vb}(\kv)$ appears at a higher screening parameter value,  $\kappa_v^{\rm L}$. From the vapor phase, we thus enter in the coexistence region. 

This method is equivalent to the usual one employed in the canonical ensemble, where phase coexistence is determined by solutions $c_{L}$ and $c_{V}$ to the simultaneous equations, $P(c_{V}) = P(c_{L})$ and $\mu_\pm(c_{V}) = \mu_\pm(c_{L})$, where $\mu_\pm = \mu_\pm^{\rm id} + \mu_\pm^{\rm ex}$ and  $P(c) = P_{\rm el}(c) + P_{\rm hc}(2c)$.

\begin{figure}[t]
\includegraphics[width=\columnwidth]{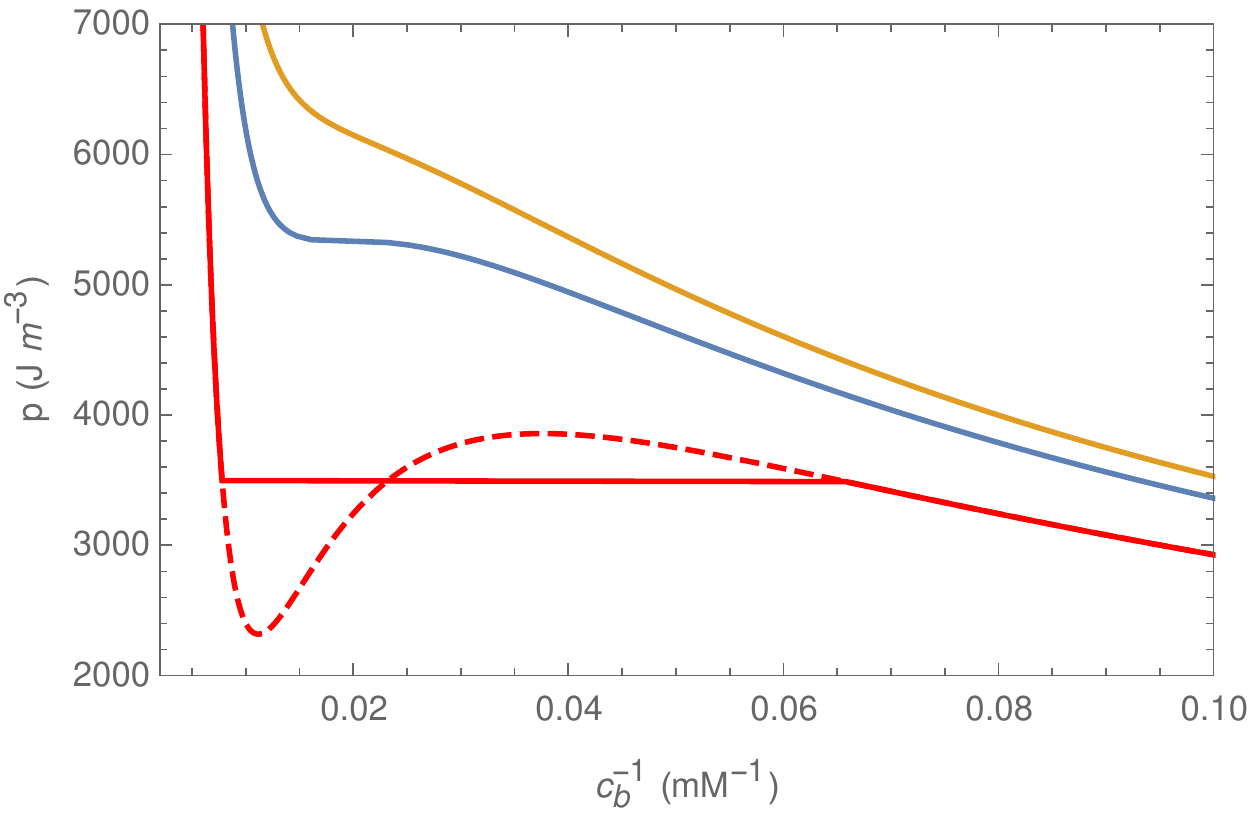}
\caption{The pressure in $\mathrm{Pa}$ as a function of the volume per ion, $c_b^{-1}$ where $c_b$ is expressed in mM, for various temperatures (from bottom to top $T= 42$, $44.16$ and $45\ \mathrm{K}$). The horizontal lines denote the beginning and the end of the phase coexistence region. The dashed lines are the pressure defined in the canonical ensemble, \eq{CSpress}, expressed in terms $c_b$. For $T=T_c=44.16 \mathrm{K}$, the phase transition is continuous.}
\label{fig:pc}
\end{figure}

\begin{figure}[t]
\includegraphics[width=\columnwidth]{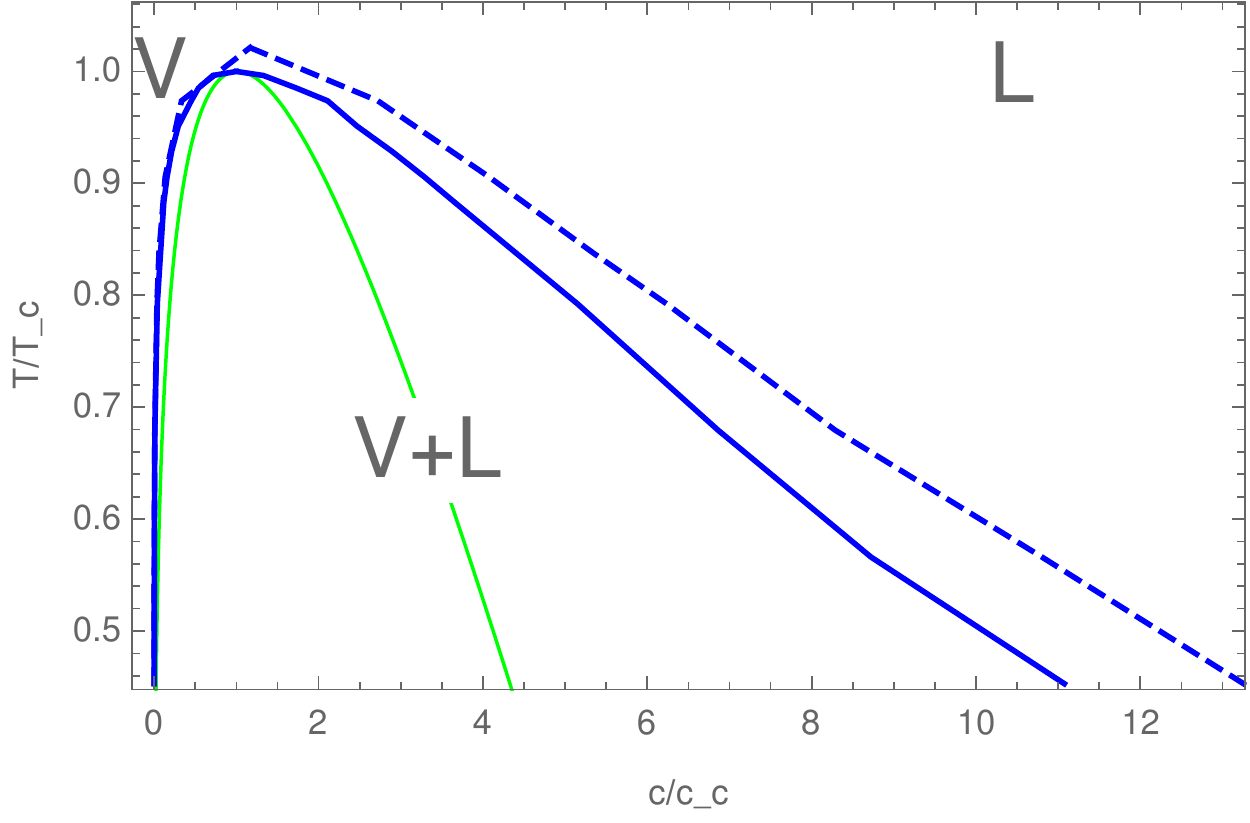}
\caption{Rescaled coexistence curve $T/T_c$ vs. $c/c_c$ in the bulk. The ``vapor'' and ``liquid'' phase are respectively on the left and right of the coexistence region. The green line is the spinodal curve and the dashed line is the coexistence curve without hardcore effects.}
\label{fig:Tcphasediagram}
\end{figure}

The total pressure $P=-w_{v}(\kb)$ is plotted with respect to the volume per ion $V/\langle N\rangle=1/c_b$ in Figure~\ref{fig:pc}. It shows the coexistence region where two solutions appear.
In Figure~\ref{fig:Tcphasediagram} is shown the coexistence region in the temperature vs. concentration plane. This figure has been plotted by identifying the two minima that appear in the variational grand potential as shown in Fig.~\ref{fig:wvb}  when varying $\lambda$ at fixed $T$ and by repeating this for various temperatures.
For our chosen value of $d$ and $\qc$ we get a critical point at $T_c=44.16$~K  and $c_c=51.91$~mM (which confirms the unphysical nature of the transition for common electrolytes).
When hardcore interactions are included in the theory, the coexistence region, as well as $T_c$ and $c_c$, are slightly reduced. The spinodal curve, $T_{\rm sp}(c)$, on which the susceptibility diverges, is defined by $(\partial P / \partial c)_{T_{\rm sp}} = 0$ and shown in green.

\begin{figure}[t]
\includegraphics[width=\columnwidth]{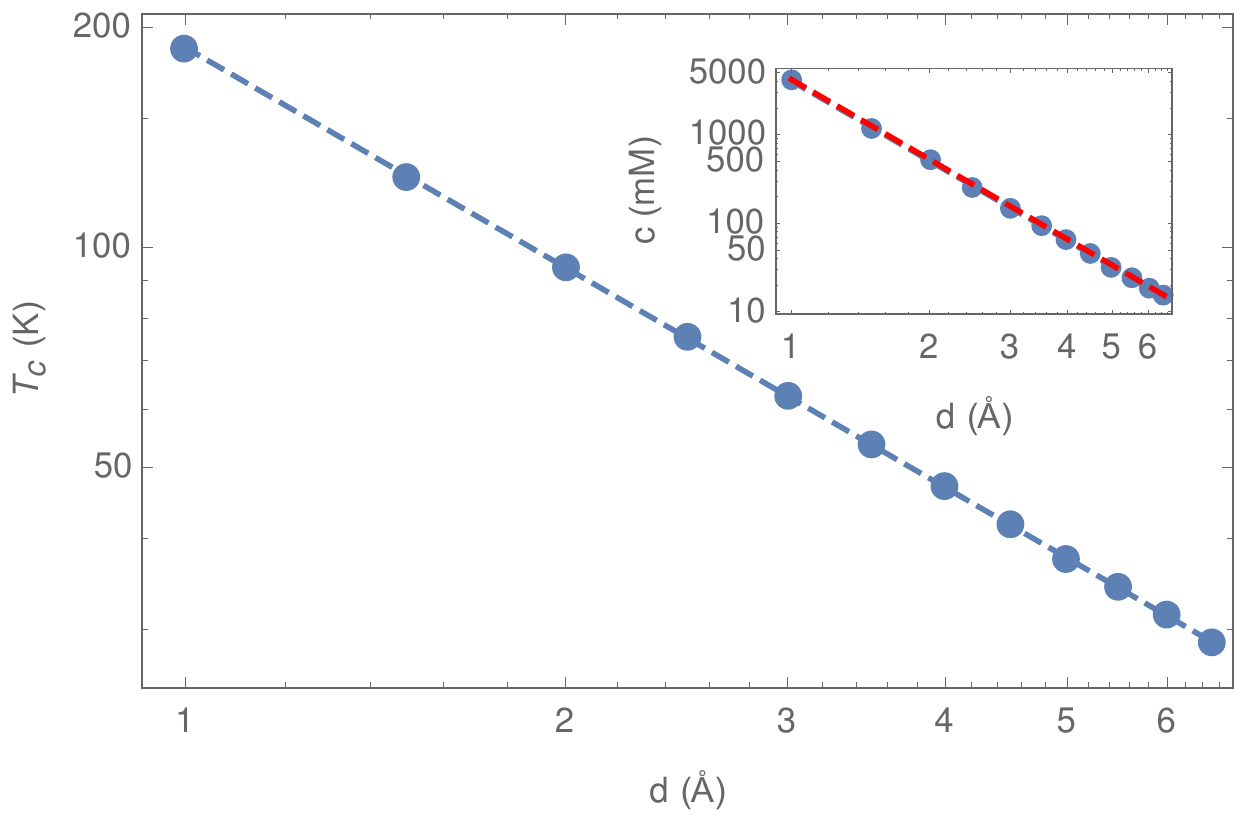}
\caption{The critical temperature as a function of the particle diameter in a log-log plot. The dots are points calculated by solving \eqs{g}{gp}. The dashed line is obtained by using the first point as a reference and assuming $T_c \propto d^{-1}$.
\textit{Inset}: The critical concentration as a function of the particle diameter in a log-log plot. The dots are points calculated by solving \eqs{g}{gp}. The blue dashed line is obtained by using the first point as a reference and assuming that $c_c \propto d^{-3}$ while the red dashed line is \eq{ccapprox}. The two lines are superimposed.}
\label{fig:Tcd}
\end{figure}

The critical temperature $T_c$ depends on the particle diameter $d$. In order to investigate this point and compare our method to other approaches, we derive the equations defining the critical point. We first note that the pressure $P$ is a function of the concentration at a fixed temperature. The critical point $(T_c,c_c)$ is defined by:
    \begin{equation}
    \left.\frac{\partial P}{\partial c}\right|_{c_c,T_c} = 0 \quad \mathrm{and} \quad \left.\frac{\partial^2 P}{\partial c^2}\right|_{c_c,T_c} = 0
    \end{equation}
After some calculation we obtain:
\begin{align}
&1 + \frac{8\eta_c - 2 \eta_c^2}{(\eta_c-1)^4}  = \frac{q^2 \kc^*}{2 \pi} \left[\arctan\left(\frac{\qc^*}{\kc^*}\right) -\frac{\kc^* \qc}{\qc^{*2} + \kc^{*2}}\right]\label{g}\\
&\frac1{2 \eta_c} - \frac{5 \eta_c^2 - 25 \eta_c -4}{(\eta_c -1)^5}  = \frac{2 q^4}{ \v^*} \frac{\qc^{*3}}{(\qc^{*2}+\kc^{*2})^2} \label{gp}
\end{align}
where we have introduced the dimensionless parameters $\eta_c = 2 \v c_c$, $\kc^* = \sqrt{8 \pi q^2 \lc^3 c_c}$, $\v^* = \v/\lc^3$, $\qc^* = \lc \qc$ and $\lc$ is the Bjerrum length at the temperature $T_c$. \eqs{g}{gp} both depend  on the cut-off, $\qc^*$, and the excluded volume $\v^*$.
Now suppose that the critical temperature is $T_{c,1}$ for a particle diameter $d_1$, which defines the hardcore parameters $\Lambda_{c,1}$ and $\v_1$. The critical temperature $T_{c,2}$ corresponding to a diameter $d_2$ with $\Lambda_{c,2}$ and $\v_2$ is such that both $\qc^*$ and $\v^*$ are unchanged. Hence $\Lambda_{c,1} /T_{c,1} = \Lambda_{c,2} /T_{c,2}$ and $\v_1 T_{c,1}^3 = \v_2 T_{c,2}^3$.
If we now use that the excluded volume is proportional to $d^3$ and that the cut-off is proportional to $d^{-1}$, then the two previous conditions are equivalent to $T_{c,1} d_1 = T_{c,2} d_2$. In other words the critical temperature is inversely proportional to $d$.
In Figure~\ref{fig:Tcd} is plotted the critical temperature as a function of the particle diameter deduced from \eqs{g}{gp} together with the $1/d$ law. The agreement is excellent.

In order to compare our result with the literature we introduce the rescaled temperature and concentration:
\begin{equation}
\label{adimpar}
T^* = \frac{4 \pi \eo \ew d k_{\rm B} T}{(q e)^2}=\frac{d}{q^2\ell_{\rm B}}\quad \mathrm{and} \quad c^* = d^3 c
\end{equation}
Various results have been found in the literature for those two quantities. For example in the Monte Carlo simulations of Ref.~\cite{Valleau1991} the authors found $T^*_c \approx 0.07$ and $c_c^* \approx 0.07$. In Ref.~\cite{Orkoulas1994} another result was found, $T^*_c \approx 0.053$ and $c_c^* \approx 0.025$, which compared well to experiments on molten salts~\cite{Kirshenbaum1962}. These values were subsequently found to be in agreement with the theory developed by Levin and Fisher \cite{Fisher1996}. They used a Bjerrum model of ion pairing  which is introduced via an explicit ``pairing reaction" between the ions~\cite{Yeh1996}. Because explicit ion-pairing terms are not included in the present variational approach, our theory recovers a phase transition which is closer to the \DH model with added hardcore interactions. The specific values of the critical parameters given in \eq{adimpar} are in our case, $T^*_c \approx 0.088$ and $c_c^* = 0.0024$. They depend on the model used to account for hardcore interactions (the Carnahan-Starling approach in our case) but are close to the ones given in Ref.~\cite{Fisher1996} using the second virial coefficient approach, $T^*_c \approx 0.061$ and $c_c^* \approx 0.0046$. Our approach is more accurate for large ionic concentrations.

We now investigate the relation between $T_c$ and $c_c$. First note that in \eq{gp} the second term can be neglected, which reduces \eq{gp} to a second order polynomial in $\eta_c$ that can be solved to give:
\be
\label{ccapprox}
c_c = \frac{\eta_c}{2 \v} \approx \frac{\qc^{*3} }{16 \pi^2 \lc^3} \left(1 -\sqrt{1-\frac{4 \pi}{q^2 \qc^*}} - \frac{2 \pi}{q^2 \qc^*} \right)
\ee
Hence $c_c$ is proportional to $T_c^3$ and inversely proportional to $d^{3}$ as it should. In the inset of Figure \ref{fig:Tcd} is plotted the calculated $c_c$ together with \eq{ccapprox}, the agreement is excellent.

\section{Electrolyte with hardcore interactions inside a nanopore}
\label{sec:nano}

\subsection{Variational grand potential}

We consider an electrolyte in an infinite cylindrical nanopore of radius $a$ in contact with a reservoir at fixed $T$ and fixed ionic fugacities $\lambda_i$. The dielectric constant of the nanopore is $\epsilon_m\ll\epsilon_w$. In the following we chose $\epsilon_m=2$. We start from the variational grand potential per unit volume derived in~\cite{Buyukdagli2010,Buyukdagli2011} without hardcore interactions, which corresponds to the first 2 terms of \eq{varpot} with $\rho_e(\vec{r})=\sigma_s\delta(r-a)$ where $\sigma_s$ is the surface charge density of the pore:
\begin{multline}
w_{v,{\rm el}} = -\sum_i \lambda_i  e^{\frac{q_i^2 \lb}{2} \kv  - q_i \pO} \left\langle e^{-\frac{q_i^2}{2} \dvO(r;\kv)} \right\rangle  + \frac{\kv^3}{24 \pi} \\
+  \frac{\kv^2}{8 \pi \lb} \int_0^1 \dd \xi \left\langle \dvO(r;\sqrt{\xi}\kv) - \dvO(r;\kv) \right\rangle + \frac{2}{a} \s \pO  \label{wvnohc}
\end{multline}
where the brackets mean an average over the nanopore volume defined as:
\begin{equation}
 \left\langle f(r) \right\rangle = \frac{2}{a^2} \int_0^a \dd r\ r f(r)
\end{equation}
and $\dvO(r;\kv)$ is the correction to the variational kernel due to the presence of the nanopore~\cite{Buyukdagli2011}
\be
v_0(\vr,\vr')=\ell_{\rm B}\frac{e^{-\kappa_v|\vr-\vr'|}}{|\vr-\vr'|}+\delta v_0(\vr,\vr';\kappa_v)
\ee
evaluated at $\vr=\vr'$, thus defined as
\be\label{deltav0}
\delta v_0(\vr;\kappa_v)=\frac{4\ell_B}{\pi}\int_0^\infty \mathrm{d}k\sideset{}{'}\sum_{m\geq0}F_m(k;\kappa_v)I_m^2(\varkappa |\vr|)
\ee
where we note $\varkappa^2=k^2+\kappa_v^2$ and the prime on the summation sign means that the term $m=0$ is multiplied by 1/2. The function $F_m$ is
\be
F_m(k;\kappa_v)=\frac{\epsilon_w \varkappa K_m(ka)K'_m(\varkappa a)-\epsilon_m k K_m(\varkappa a)K'_m(k a)}
{\epsilon_m k I_m(\varkappa a)K'_m(k a)-\epsilon_w \varkappa K_m(k a)I'_m(\varkappa a)}
\ee
where $K_m(x)$ and $I_m(x)$ are modified Bessel functions.

We make the reasonable simplifying assumption that the dielectric exclusion near the nanopore surface is strong enough to keep finite size ions from approaching the pore wall (in fact there should be a distance of closest approach given by the ionic radius). This assumption leads to the absence of purely steric exclusion effects, which would become important for neutral particles, very large ions or inverted dielectric profiles (with a larger dielectric constant outside the nanopore than within, leading to ion accumulation at the nanopore surface~\cite{Lue2015}).

Because of the rotational symmetry around the cylinder axis and the translational symmetry along the cylinder axis, all quantities depend only on the radial distance $r$.
The first term in \eq{wvnohc} is minus the pressure of an ideal solution for which the fugacity has been modified by the electrostatic potential $\phi_0$ and the excess electrostatic chemical potential, $\mu_{{\rm el},i}^{\rm ex}$ given in \eq{muexnohc}. The second term is the \DH electrostatic contribution which is also present in the bulk phase. The last two terms are the electrostatic contributions due to the presence of the nanopore.

We follow the same strategy as above by replacing the first term in \eq{wvnohc} by minus the pressure of the Carnahan-Starling approach, \eqs{CSpress}{relfugacity}, with the modified fugacity given in \eq{wvnohc}.
We then replace the second term by the equivalent terms of \eq{bulkgrdpot} including the short range cut-off $\qc$ and we leave the last terms identical. The resulting variational grand potential is:
\bea
 w_{v} &= &
     \delta w_{vb} \label{wv}\\
     & - & P_{\rm hc} \left( \sum_i \lambda_i  e^{\frac{q_i^2\lb}{\pi} \kv \at{\frac{\qc}{\kv} } - q_i \pO} \left\langle e^{-\frac{q_i^2}{2} \dvO(r;\kv)} \right\rangle \right) \nonumber \\
     & + & \frac{\kv^2}{8 \pi \lb} \int_0^1 \dd \xi \left\langle \dvO(r;\sqrt{\xi}\kv) - \dvO(r;\kv) \right\rangle + \frac{2}{a} \s \pO   \nonumber
\eea
where the function $P_{\rm hc}(\bar \lambda)$ is given in \eq{CSpress}. This approach thus conserves some important properties of the system, as we will see in the next section. For our purposes here the simple ``average density" approach (in Density Functional Theory terminology) to inhomogeneous systems embodied in the above choice  for the hardcore contribution to \eq{wv} suffices.
One can check that in the limits of large cutoff $\qc/\kv \gg 1$ and small packing fractions $\eta \ll 1$, \eq{wv} and \eq{wvnohc} are equivalent.

The variational equation for $\pO$ is simply:
\begin{equation}
\frac{2 \s}{a} + \sum_i q_i c_i = 0
\label{phi0}
\end{equation}
where the concentration inside the pore of ionic species $i$ is given by:
\begin{equation}
 c_i = \frac{1}{\v}\eta(\bar \lambda) \frac{\lambda_i  e^{\frac{q_i^2\lb}{\pi} \kv  \at{\frac{\qc}{\kv} } - q_i \pO} \left\langle e^{-\frac{q_i^2}{2} \dvO(r;\kv)} \right\rangle}{\sum_j \lambda_j  e^{\frac{q_j^2\lb}{\pi} \kv  \at{\frac{\qc}{\kv} } - q_j \pO} \left\langle e^{-\frac{q_j^2}{2} \dvO(r;\kv)} \right\rangle}
\label{cihc}
\end{equation}
where $\eta(\bar \lambda)$ is solution of \eq{relfugacity} and $\bar \lambda$ is the argument of $P_{\rm hc}$ in \eq{wv}. We therefore recover $\eta = \v \sum_i c_i$, the packing fraction inside the cylinder, and \eq{phi0} is simply the condition of charge conservation.

The variational equation for $\kv$ is more involved. The details of the calculation are reported in the Appendix~\ref{appB}, and one obtains
\begin{multline}
\label{simplifvareq}
\frac{\kv^2}{4\pi \ell_{\rm B}} \left[\frac{2\ell_{\rm B}}{\pi} \at{\frac{\qc}{\kv}} -  \frac{2\ell_{\rm B}}{\pi} \frac{\kv \qc}{\kv^2 + \qc^2} - \left\langle  \dvOp(r;\kv)   \right\rangle \right] \\
= \sum_i q_i^2 c_i  \left[ \frac{2\ell_{\rm B}}{\pi}  \at{\frac{\qc}{\kv}} - \frac{2\ell_{\rm B}}{\pi}  \frac{\kv \qc}{\kv^2 + \qc^2} \phantom{\frac{ \left\langle \dvOp(r;\kv) e^{-\frac{q_i^2}{2} \dvO(r;\kv)} \right\rangle}{\left\langle e^{-\frac{q_i^2}{2} \dvO(r;\kv)} \right\rangle}} \right. \\
\left. - \frac{ \left\langle \dvOp(r;\kv) e^{-\frac{q_i^2}{2} \dvO(r;\kv)} \right\rangle}{\left\langle e^{-\frac{q_i^2}{2} \dvO(r;\kv)} \right\rangle}  \right]
\end{multline}
which gives the modified \DH relation for $\kv$ in the nanopore. Clearly $\kv$ vanishes for neutral particles, $q_i \rightarrow 0$. Note that the contribution of the hardcore excess chemical potential enters implicitly though the expression of $c_i$.

\subsection{Partition coefficients and phase diagram}

The Potential of Mean Force (PMF), $\pmf_i$, and the partition coefficient $k_i$ are defined as:
\begin{multline}
k_i  \equiv \frac{c_i}{c_{b,i}} = \left\langle e^{-\pmf_i(r;\kv)} \right\rangle \\
= e^{\frac{q_i^2\lb}{\pi}\left[ \kv \at{\frac{\qc}{\kv} } - \kb \at{\frac{\qc}{\kb}}\right] - q_i \pO} \\ \left\langle  e^{-\frac{q_i^2}{2} \dvO(r;\kv)} \right\rangle e^{\frac{\nb -3}{(\nb -1)^3} - \frac{\eta -3}{(\eta-1)^3}}
 \end{multline}
 where $c_i$ and $c_{b,i}$ are the concentrations of ion $i$ in the pore and in the bulk, respectively.
We therefore define an \textit{effective} PMF as:
\begin{multline}
 \pmf_i (r;\kv) = \frac{q_i^2}2 w_{\rm el}(r;\kappa_v)+ q_i\phi_0+w_{\rm hc}\\
 =-\frac{q_i^2 \lb}{\pi}\left[ \kv \at{\frac{\qc}{\kv} } - \kb \at{\frac{\qc}{\kb} }\right]  \\
  +\frac{q_i^2}{2} \dvO(r;\kv)+ q_i \pO - \frac{\nb -3}{(\nb -1)^3} + \frac{\eta -3}{(\eta-1)^3}
\end{multline}
where the dependence on $r$ comes only from the term containing $\dvO(r;\kv)$, the other spatial dependencies being integrated out.
This effective PMF has 3 contributions, the electrostatic one $w_{\rm el}$ associated to the kernel $v_0$, the elecrostatic energy $q_i\phi_0$, and the hardcore contribution $w_{\rm hc}$. It allows us to simplify the variational equation \eq{simplifvareq} as
\be
\kappa_v^2=\frac{4\pi\ell_{\rm B}}{\left\langle \frac{\partial w_{\rm el}(r;\kappa_v)}{\partial \kappa_v}  \right\rangle}\sum_i q_i^2 c_{b,i}\left\langle \frac{\partial w_{\rm el}(r;\kappa_v)}{\partial \kappa_v} e^{-\pmf_i(r;\kv)} \right\rangle.
\ee
Therefore the direct hardcore interactions associated to the Carnahan-Starling pressure enter through $w_{\rm hc}$ only in the Boltzmann factor. In the limit $\eta\to0$ and $\Lambda_c\to \infty$, we recover the variational equation of Ref.~\cite{Buyukdagli2011}.

\begin{figure}[t]
\includegraphics[width=\columnwidth]{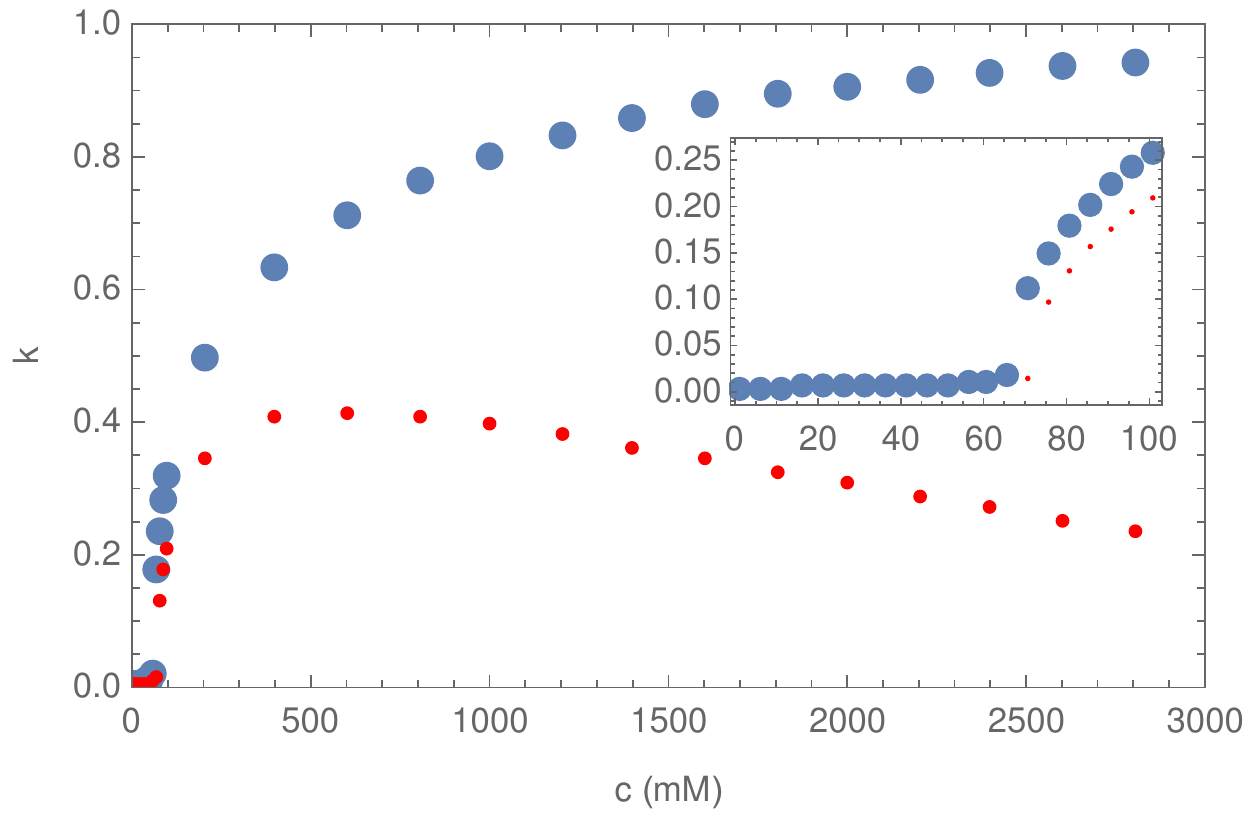}\\
\includegraphics[width=\columnwidth]{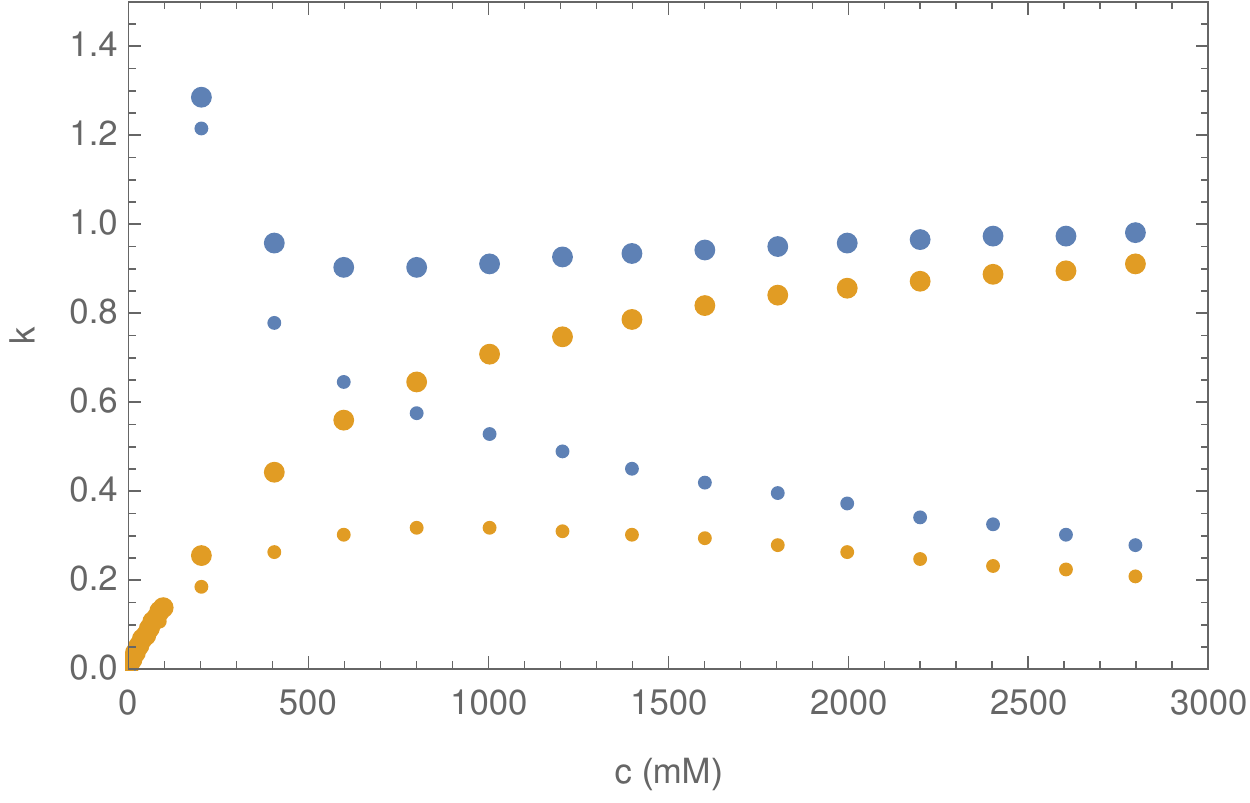}
\caption{The partition coefficient as a function of the bulk concentration  ($a=1$~nm). The big circles are with excluded volume effect while the small circles are without. {\it Top}: Neutral pore ($\s = 0$), the inset is a zoom on the first 100 mM. {\it Bottom}: Charged pore ($\s = -0.01\ e/\mathrm{nm^2}$) with coions in yellow (bottom) and counterions in blue (top).}
\label{fig:kneutral}
\end{figure}

In Figure~\ref{fig:kneutral} are plotted the partition coefficients of a symmetric electrolyte in a neutral and a charged pore.
In these figures, we compare the results obtained from the variational grand potential without the hardcore interactions, \eq{wvnohc}, (small dots) and the ones with the hardcore interactions, \eq{wv} (large dots).
We see that, for a high bulk ionic concentration, $c_b$, the partition coefficients without hardcore interactions decrease with $c_b$.
This unphysical result is an artifact of the approach, because the ions tends to form neutral pairs with ions of opposite sign on top of each other, as the concentration increases. Of course, this does not happen when hardcore interactions are included in the model, the partition coefficients now slowly saturate to 1 as $c_b$ increases.
For a low enough $c_b$, the two approaches lead to the same partition coefficients, although the difference between the two occurs for $c_b > 100$~mM. For a neutral pore (Fig.~\ref{fig:kneutral} Top), $k$ is a monotonously increasing function of $c_b$ and identical for anions and cations. For a charged pore with surface charge density $\s = -0.01\ \mathrm{e/nm^2}$ (Fig.~\ref{fig:kneutral} Bottom), coions have an increasing partition coefficient, $k_-$, which evolves similarly to the case of neutral pore. For counterions, however, $k_+>1$ for low $c_b$ because $c_+$ is controlled by the surface charge density and then decreases down to $k_+\approx0.9$ for $c_b\approx0.5$~M. For higher $c_b$, $k_+$ increases slowly up to 1, so that $c_+\simeq c_b$. 
%

\begin{figure}
\includegraphics[width=\columnwidth]{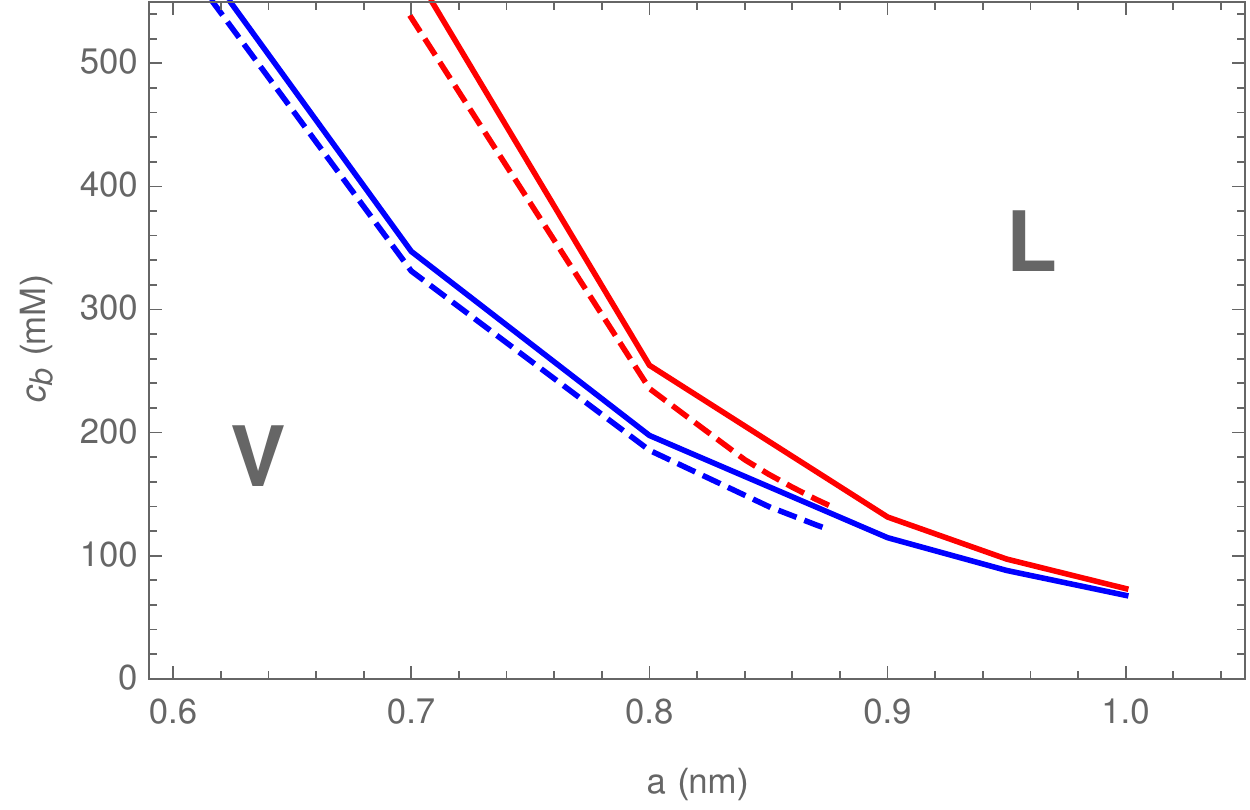}
\caption{Ionic liquid-vapor phase diagram in a nanopore in the bulk concentration ($c_{b}$)--nanopore radius ($a$) plane. The critical lines separates the ionic exclusion ``vapor'' state (V)  from the ionic penetration ``liquid'' state (L). The blue lines include the hardcore effect while the red lines does not. The solid lines correspond to $\s = 0$ and the dashed lines to $\s = 10^{-3}\ e/\mathrm{nm^2}$.}
\label{fig:diagram}
\end{figure}

In the inset of Figure~\ref{fig:kneutral} Top, we see the signature of the phase transition studied in Ref.~\cite{Buyukdagli2011}. It is a first order phase transition between a phase where ions are excluded from the nanopore, the so called ionic ``vapor'' phase, and a phase where ions enter the pore, the ionic ``liquid'' phase.
This transition exists at room temperature for small enough pore radii. At a critical radius $a_c$, the phase transition becomes continuous and then disappears for $a>a_c$. Being a room temperature transition, the complications due to strong ion pairing and clustering that occur in bulk electrolytes (and lead to quantitative disparities between the present approach and MC simulations), may perhaps be minimized (it would thus be of great interest to carry out MC simulations in a nanopore to detect the predicted transition).
In Figure~\ref{fig:diagram} is shown the phase diagram obtained with (blue lines) and without (red lines) hardcore interactions and with (dashed lines) and without (solid lines) a small charge density on the nanopore. The coexistence lines without hardcore interactions are the same as the ones obtained in Ref.~\cite{Buyukdagli2011}.
For a given small nanopore radius $a$, hardcore interactions decrease the value of the bulk concentration at which the transition takes place. However for radii close to $a_c$, the difference between the two approaches decreases and the critical point is unchanged.
The effect of a non zero surface charge density $\s$ is to further decrease the value of the critical radius $a_c$ and to increase the bulk ionic concentration $c_{b,c}$ at the transition. The shapes of the coexistence lines are, however, very similar.
For sufficiently large surface charge densities, the phase transition disappears.

\section{Conclusion}
\label{sec:concl}

By introducing the Carnahan-Starling pressure and a cut-off in Fourier space for the Debye-H\"uckel contribution to the grand potential, we develop a rigorous variational approach that includes ion-ion correlations modified by the dielectric jump and hardcore interactions. We consider both a bulk electrolyte and an electrolyte confined in a nanopore. First, we are able to recover important features of the bulk restricted primitive model, such as the increase of the excess chemical potential for large concentrations and the liquid-vapor phase transition induced by ion-ion correlations.
This approach allows us to study not only the behavior of charged hard spheres in a bulk phase, but also in the more complex case of a neutral or charged cylindrical nanopore. In the latter case the phase transition is induced by the dielectric exclusion and therefore occurs at room temperature for small pore radii (and ions sizes corresponding to those of common mineral salts). The ionic partition function is obtained for a whole range of reservoir concentrations, from very low ones up to saturation $c_b\simeq 3$~mol/L. 

Possible extensions of our theory include the use of pressure expression taking into account the different sizes of the ions, as developed in Ref.~\cite{Lebowitz1964,Mansoori1971} for neutral liquids, the use of a more accurate variational kernel with a spatially dependent variational Debye screening parameter, $\kappa_v(\vec{r})$ and the integration of a more sophisticated approach to inhomogeneous liquids (such as local density approximations).
One weakness of our approach is that it gives a critical concentration for the phase transition which is too low compared to bulk MC simulations. This is probably because we did not consider explicit ion pairing in the theory.
Finally, our approach yields a theoretical framework for computing the transport coefficients for electrolytes in a single well characterized nanopores, experiments which are now accessible~\cite{Balme2015}.

\acknowledgments

Financial support by the French Research Program ANR-BLANC (TRANSION project, ANR-2012-BS08-0023) is acknowledged. We are tributary to the Centre National de la Recherche Scientifique (CNRS) and the Universities of Toulouse III-Paul Sabatier and Montpellier.

\appendix

\section{Variational approach}
\label{appA}

We start from the grand partition function of hardcore particles in an external potential \eq{grandpotHCfield}~\cite{Negele1992}, and we introduce the electrostatic interaction for point particles which interact with the Coulomb potential \eq{Coulomb}. After performing a Hubbard-Stratonovitch transformation and introducing a fluctuating field $\phi(\vr)$, the electrostatic grand canonical partition function of an electrolyte is
\begin{multline}
\label{grandpotcoul}
 \Xi_{\rm el}[\mu_\gamma] = \frac{1}{Z_{\rm C}}\int  \DD \phi\,  \exp
 \left[
 -H_{\rm c}[\phi(\vec{r})] + i \int_{\vec{r}} \rho_{\rm e}(\vec{r})\phi(\vec{r})\right.\\
 \left.+ \sum_\alpha \frac{1}{V_{\alpha}} e^{ \frac{1}{2} q^2_\alpha v_{\rm c}(0) } \int_{\vec{r}} g(\vec{r})
 e^{ \mu_\alpha + i q_\alpha \phi(\vec{r}) }
 \right],
 \end{multline}
 where $\rho_{\rm e}(\vec{r})$ is the external fixed charge density (in units of $e$) and
\be
 Z_{\rm C} = \int \mathcal{D}\phi\; e^{-H_{\rm c}[\phi(\vec{r})]} = -\frac12\mathrm{tr}\, \ln (v_{\rm c})
\ee
with $H_{\rm c}$ defined in \eq{Hc0}.
The exact average electrostatic potential, which obeys the Poisson equation with both the ion charge density and external charge density in the source term, is given by $\Phi(\vr) = -i\langle\phi(\vr)\rangle$.

If we now combine the electrostatic interactions with the hardcore ones, the grand partition function becomes~\cite{Netz1999}:
\be
 \Xi_{\rm el,hc} = \frac{1}{Z_U}\int \prod_\nu \DD \psi_\nu \,
 e^{ -H_{\rm hc}[\psi_\gamma(\vec{r})] } \Xi_{\rm el}[\mu_\gamma +i\psi_\gamma(\vec{r})].
\ee
Performing the functional integral over the hardcore fields in the absence of an external potential yields an effective electrostatic problem~\cite{Lue1999a}:
\begin{multline}
\label{grandpartelecfield}
 \Xi_{\rm el,hc}[\mu_\gamma]  = \frac{1}{Z_c} \int  \DD \phi\,
 e^{ -H_{\rm c}[\phi(\vec{r})] }
 \exp \left\{ \int i \rho_{\rm e}(\vec{r})\phi(\vec{r}) \right.\\
 \left. -
  \Omega_{\rm hc}\left[ \mu_\alpha+\frac12 q^2_\alpha v_{\rm c}(0)+ i q_\alpha \phi(\vec{r})\right] \right\},
\end{multline}
where $\Omega_{\rm hc}$ is given in \eq{omegahc}. Eq.~(\ref{grandpartelecfield}) cannot be evaluated exactly, but has been treated in mean-field theory~\cite{Lue1999a}. In the limit of vanishing hardcore interactions $\Omega_{\rm hc}$ reduces to the ideal gas form:
\begin{multline}
\Omega_{\rm id}\left[ \mu_\alpha+\frac{1}{2} q^2_\alpha v_{\rm c}(0)+ i q_\alpha \phi(\vec{r})\right] = \\ - \sum_\alpha \frac{1}{V_{\alpha}} e^{ \frac{1}{2} q^2_\alpha v_{\rm c}(0) } \int_{\vec{r}} g(\vec{r})
 e^{ \mu_\alpha + i q_\alpha \phi(\vec{r}) },
 \end{multline}
and we recover Eq.\eqref{grandpotcoul}.

Although evaluating the electrostatic part of the grand partition function at fixed $\psi_\gamma(\vec{r})$,
\begin{multline}
 \Xi_{\rm el}[\mu_\gamma + i\psi_\gamma(\vec{r})] = \frac{1}{Z_c}\int  \DD \phi\,
 e^{ -H_{\rm c}[\phi(\vec{r})] }
 \exp
 \left[
  \int_\vec{r} i \rho_{\rm e}\phi \right. \\
  \left. + \sum_\alpha \frac{1}{V_{\alpha}} e^{\frac{1}{2} U_{\alpha\alpha}(0) + \frac{1}{2} q^2_\alpha v_{\rm c}(0)} \int_\vec{r} \, g(\vec{r})  e^{i \psi_\alpha(\vec{r}) +  \mu_\alpha + i q_\alpha \phi(\vec{r}) }
 \right],
\end{multline}
is also intractable due to non-linear terms,  we can use a Gaussian variational method that consists in introducing a variational Gaussian Hamiltonian defined in \eq{HOdef} and then rewriting $\Xi_{\rm el}[\mu_\gamma + i\psi_\gamma(\vec{r})]$ as
\begin{multline}
 \Xi_{\rm el}[\mu_\gamma + i\psi_\gamma(\vec{r})]
  = \frac{Z_0}{Z_c}
 \left\langle
 \exp
 \left[
 H_0[\phi] - H_{\rm c}[\phi] +
  \int_\vec{r} i \rho_{\rm e}\phi \right. \right. \\ \left. \left.
  + \sum_\alpha \frac{1}{V_{\alpha}} e^{\frac12 U_{\alpha\alpha}(0) + \frac12 q^2_\alpha v_{\rm c}(0)} \int_\vec{r} \, g(\vec{r})  e^{i \psi_\alpha(\vec{r}) +  \mu_\alpha + i q_\alpha \phi(\vec{r}) }
 \right]
 \right\rangle_0
 \end{multline}
where the expectation value is evaluated with the  variational Gaussian Hamiltonian $H_0[\phi]$.
If we introduce a variational electrostatic grand partition function,
\begin{multline}
 \Xi_{\rm el}^{\rm v}[\mu_\gamma + i\psi_\gamma(\vec{r})]
  = \frac{Z_0}{Z_c}
 \exp
 \left[
 \left\langle
 H_0[\phi] - H_{\rm c}[\phi] +
  \int_\vec{r} i \rho_{\rm e}\phi\right. \right. \\ \left. \left.
  + \sum_\alpha \frac{1}{V_{\alpha}} e^{\frac{1}{2} U_{\alpha\alpha}(0) + \frac{1}{2} q^2_\alpha v_{\rm c}(0)} \int_\vec{r} \, g(\vec{r})  e^{i \psi_\alpha(\vec{r}) +  \mu_\alpha + i q_\alpha \phi(\vec{r}) }
 \right\rangle_0
 \right] \\
 = \frac{Z_0}{Z_c}
 \exp
 \left[
 \left\langle
 H_0[\phi] - H_{\rm c}[\phi]
  \right\rangle_0 -
  \int_\vec{r}  \rho_{\rm e}\phi_0\right.\\
\left.  + \sum_\alpha \frac{1}{V_{\alpha}} e^{\frac{1}{2} U_{\alpha\alpha}(0)} \int_\vec{r} \, g(\vec{r})  e^{i \psi_\alpha(\vec{r}) +  \mu_\alpha - u_\alpha^0(\vec{r}) }
 \right]
 \end{multline}
 where $\phi_0(\vec{r}) = -i \langle  \phi(\vec{r})  \rangle_0  $
 and
 \be
 u_\alpha^0(\vec{r}) = q_\alpha \phi_0(\vec{r}) +  \frac{1}{2} q^2_\alpha
 \left[ v_0(\vr, \vr) - v_{\rm c}(0)\right],
 \ee
 then by the Gibbs-Bogoliubov-Feynman inequality:
 \be
 \Xi_{\rm el}^{\rm v}[\mu_\gamma + i\psi_\gamma(\vec{r})]\leq \Xi_{\rm el}[\mu_\gamma + i\psi_\gamma(\vec{r})].
 \ee
 Therefore
 \begin{multline}
\label{grandparttotalfield}
 \Xi_{\rm el,hc} = \frac{1}{Z_U}\int \prod_\nu \DD \psi_\nu \,
 e^{ -H_{\rm hc}[\psi_\gamma] } \Xi_{\rm el}[\mu_\gamma + i\psi_\gamma(\vec{r})] \\
    \geq \Xi_{\rm el,hc}^{\rm v}= \frac{1}{Z_U}\int \prod_\nu \DD \psi_\nu \,
 e^{ -H_{\rm hc}[\psi_\gamma] } \Xi_{\rm el}^{\rm v}[\mu_\gamma + i\psi_\gamma(\vec{r})],
\end{multline}
 which implies that  the variational grand-potential \eq{varpot} is an upper bound to the exact grand-potential $\Omega_{\rm el,hc} = - \ln (\Xi_{\rm el,hc}) \leq \Omega_{v}$ with $\Omega_0=-\frac12\mathrm{tr}\, \ln(v_0/v_{\rm c})$.

\section{Variational equation in nanopore}
\label{appB}

By minimizing \eq{wv} with respect to $\kappa_v$, one finds
\begin{align}
 0&= \frac{\kv^2}{4 \pi^2} \left[ \at{\frac{\qc}{\kv}} - \frac{\kv \qc}{\kv^2 + \qc^2}\right]\label{notsimple}  \\
 &+ \frac{\kv}{4 \pi \lb}  \int_0^1 \dd \xi \left\langle \dvO(r;\sqrt{\xi}\kv) -  \dvO(r;\kv) \right\rangle \nonumber \\
 &+ \frac{\kv^2}{8 \pi}  \int_0^1 \dd \xi \left\langle \sqrt{\xi} \dvOp(r;\sqrt{\xi}\kv) -  \dvOp(r;\kv) \right\rangle \nonumber \\
 & - \frac{\partial P_{\rm hc}}{\partial \kv}\left(  \sum_i \lambda_i  e^{\frac{q_i^2}{\pi} \kv \lb \at{\frac{\qc}{\kv} } - q_i \pO} \left\langle e^{-\frac{q_i^2}{2} \dvO(r;\kv)} \right\rangle  \right)\nonumber
 \end{align}
Noting that
\begin{equation}
\label{partialP}
 \frac{\partial P_{\rm hc}}{\partial \kv} = \frac{\eta}{v \bar \lambda} \frac{\partial \bar \lambda}{\partial \kv}
\end{equation}
where
\begin{equation}
 \label{Xdef}
\bar \lambda =  \sum_i \lambda_i  e^{\frac{q_i^2}{\pi} \kv \lb \at{\frac{\qc}{\kv} } - q_i \pO} \left\langle e^{-\frac{q_i^2}{2} \dvO(r;\kv)} \right\rangle
\end{equation}
and therefore
\bea
 \frac{\partial \bar \lambda}{\partial \kv} &=&\frac12\sum_i \lambda_i q_i^2  e^{\frac{q_i^2}{\pi} \kv \at{\frac{\qc}{\kv} } - q_i \pO} \label{partialX}\\
 &\times&  \left\{\frac2{\pi}\left[\at{\frac{\qc}{\kv}} - \frac{\kv \qc}{\kv^2 + \qc^2}  \right]\left\langle e^{-\frac{q_i^2}{2} \dvO(r;\kv)} \right\rangle  \right. \nonumber \\
& &\left. - \left\langle  \dvOp(r;\kv) e^{-\frac{q_i^2}{2} \dvO(r;\kv)} \right\rangle\right\}, \nonumber
\eea
\eq{notsimple} can be simplified by using the following identity:
\begin{multline}
\int_0^1 \dd \xi \left\langle \sqrt{\xi} \dvOp(r;\sqrt{\xi}\kv) \right\rangle =\\
\frac2{\kv} \int_0^1 \dd \xi \left\langle  \dvO(r;\kv)-\dvO(r;\sqrt{\xi}\kv) \right\rangle.
\label{intxi}
\end{multline}
Putting together Eqs.\eqref{cihc}, \eqref{partialP}, \eqref{partialX} and \eqref{intxi}, we rewrite \eq{notsimple} as \eq{simplifvareq}.

\bibliography{variational}

\end{document}